\documentclass[aps,prl,twocolumn,groupedaddress,showpacs]{revtex4-1}

\usepackage{graphicx}
\usepackage{color}
\usepackage{amsmath}
\usepackage[T1]{fontenc} % Use modern font encodings

\newcommand{\M}{{\bf M}}
\newcommand{\m}{{\bf m}}

\begin{document}

\title{Accelerating, guiding, and compressing skyrmions by defect rails}

\author{Josep Castell-Queralt}
\author{Leonardo G. Gonz\'alez-G\'omez}
\author{Nuria Del-Valle}
\author{Alvaro Sanchez}
\author{Carles Navau}\email{carles.navau@uab.cat}
\affiliation{Departament de F\'{\i}sica, Universitat Aut\`onoma de Barcelona, 08193 Bellaterra, Barcelona, Catalonia, Spain}

\begin{abstract}
Magnetic skyrmions are promising candidates as information carriers in spintronic devices. The transport of individual skyrmions in a fast and controlled way is a key issue in this field. Here we introduce a novel platform for accelerating, guiding and compressing skyrmions along predefined paths. The guiding mechanism is based on two parallel line defects (rails), one attractive and the other repulsive. Numerical simulations, using parameters from state-of-the-art experiments, show that the speed of the skyrmions along the rails can be increased up to an order of magnitude  with respect to the non-defect case whereas the distance between rails can be as small as the initial radius of the skyrmions. In this way, the flux of information that can be coded and transported with magnetic skyrmions could be significantly increased.
\end{abstract}

\maketitle

\section{Introduction}
Magnetic skyrmions are being proposed as information carriers for a new generation of ultradense magnetic memories and logic devices \cite{Fert2013}.
In these systems, the presence or absence of a skyrmion usually codes information bits. Due to their topological protection, relatively small deformations in the skyrmionic structure do not destroy the information. This fact, together with their small size (few tens of nanometers), can promote them as future building blocks for high-density magnetic transport and storage systems.

\begin{figure}[!b]
	\centering
		\includegraphics[width=0.45\textwidth]{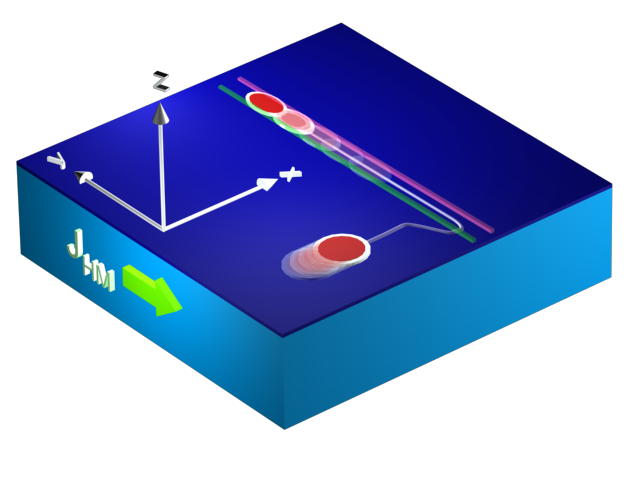}
	\caption{Sketch of the proposed skyrmionic rails. Two line defects in the ferromagnet (dark-blue thin film) can guide the skyrmion, which is driven by the torque produced by spin-polarized current generated due to the spin-Hall effect in the heavy metal (light-blue) after applying a current-density ${\bf J}_{HM}$. When the skyrmion is channeled between the two defects, its speed can be increased.}
	\label{fig:sketch}
\end{figure}

Racetracks with spin-polarized currents have been proposed and used to transport skyrmions \cite{Sampaio2013,Tomasello2014,Iwasaki2013NatNano,Fert2013,Woo2016}. In these systems, racetrack borders create a confining potential \cite{Navau2016,Chen2017,Song2017,Chen2017,Zhang2015NJP,Muller2017} that helps the movement of skyrmions along the track for electronic-spin densities above a given threshold, and avoiding the skyrmionic Hall effect because of the borders repulsion. Confinement of skyrmions along a track has also been achieved by manipulating the anisotropy \cite{Purnama2015,Song2017,Upadhyaya2015,Kang2016}, the Dzyaloshinskii-Moriya interaction \cite{Mulkers2017}, or by adding an in-plane magnetized boundary layer \cite{Zhang2017Nanoscale}. The parallelization of transport using multichannel or several lanes for skyrmions \cite{Muller2017,Song2017} could increase the flux of information that can be carried for a given total width. However, although the presence of one border can increase the velocity of the skyrmions close to that border \cite{Navau2016,Iwasaki2014}, their confinement between two borders can produce a significant change in the skyrmion size and reduce their velocity. In particular, as shown in Ref. \cite{Chen2017}, when skyrmions move along width-narrowing nanotracks their size and speed are reduced.

Here we present a novel concept for transporting skyrmions in films along a defined lane which guides the skyrmion while producing a significant increase in the skyrmionic speed as well. The lane can be as small as the radius of the original skyrmion, thus enhancing the flux of information that can be transported for a given width of material. To achieve this goal, we take profit of the tunable effect that artificial defects can produce over skyrmions \cite{Stosic2017,Kim2017PRB,Navau2018,Muller2015,Liu2013,LimaFernandes2018}. The central idea is to guide skyrmions along a lane defined by two parallel line defects (rails), tailored in such a way that one of them is attractive and the other repulsive. With the adequate conditions, the skyrmion is channeled between the two line defects and its speed is increased because the two rails thrust it in the same direction (sketch in Fig. \ref{fig:sketch}).

\section{Model}
We model the system by considering a planar ($xy$-plane) ultrathin ferromagnetic film on top of a  non-magnetic heavy metal (HM) substrate with strong spin-orbit coupling, which provides interfacial Dzyaloshinskii-Moriya (iDM) interaction between the magnetic moments in the ferromagnet. The micromagnetic framework is used, where the magnetization distribution $\M$ has constant modulus $M_s$ (the saturation magnetization) all over the ferromagnetic film. Once a skyrmion is created in the film, an electronic current flowing in the HM with density ${\bf J}_{HM}$ will result, by means of spin-Hall effect (SHE), in a spin-polarized current able to drive that skyrmion. The torque generated over the magnetization is given by \cite{Liu2011,Song2017,Finocchio2013,Knoester2014}:
\begin{equation}
{\bf T}_{SHE}= -\frac{\mu_B \theta_H J_{HM}}{e M^2_s d} \M \times \M \times \hat{\bf \sigma},
\label{eq.SHETorque}
\end{equation}
where $\mu_B$ is the Bohr magneton, $\theta_H$ is the Hall angle, and ${\bf \hat{\sigma}}$ the direction of the polarization of the spin-polarized electrons flowing into the ferromagnet. $e$  ($>0$) is the charge of the electron and $d$ is the thickness of the ferromagnetic layer. 

The dynamics of the system will be simulated by solving the micromagnetic Landau-Lifshitz-Gilbert equation \cite{Zhang2004} using a specific homemade code. In the dimensionless version, this equation becomes
\begin{equation}
\frac{{\rm d} {\bf m}}{{\rm d} \tau}  = - (\m \times {\bf h}_{eff}) + \alpha \m \times \frac{{\rm d}  {\bf m}}{{\rm d} \tau} + {\bf t}_{SHE},
\label{eq.LLG}
\end{equation}
where $\alpha$ is the Gilbert damping constant, $\tau$ is the time normalized to $t_0=1/(\gamma M_s)$, i. e. $\tau=t/t_0$, $\gamma$ is the gyromagnetic constant ($\gamma=2.21\cdot 10^{5}$ m A$^{-1}$ s$^{-1}$), ${\bf t}_{SHE}$ is the normalized torque ${\bf t}_{SHE}=(t_0/M_s){\bf T}_{SHE} $, ${\bf m}$ is the normalized magnetization ${\bf m}=\M/M_s$, and ${\bf h}_{eff}$ is normalized (to $M_s$) effective field that includes exchange, iDM, and uniaxial perpendicular anisotropy interactions,
\begin{equation}
{\bf h}_{eff}=\nabla^2 \m + \xi [(\nabla \cdot \m){\bf \hat{z}}-\nabla m_z] + \kappa m_z {\bf \hat{z}}.
\end{equation}
All the length dimensions have been normalized to the exchange length $l_{ex}=\sqrt{2A/\mu_0 M_s^2}$ ($A$ is the exchange constant). $\xi=D l_{ex}/A$ ($D$ is the iDM constant) and $\kappa=2K/\mu_0M_s^2$ ($K$ is the uniaxial anisotropy constant) are dimensionless variables. Because ultrathin films are being considered, the demagnetizing fields can be taken into account using an effective anisotropy constant \cite{Mulkers2017,Coey2010}. 

In our calculations we use typical parameters for a Co/Pt/AlO$_x$ ultrathin film (thickness $d=0.6$nm) system with strong iDM interaction consistent with experiments \cite{Sampaio2013,Raposo2017,Metaxas2007,Fert2013,Woo2016,Del-Valle2015,Chen2017,Schellekens2013,Belmeguenai2015,Kim2017APL}: $M_s = 580$ kA m$^{-1}$, $A=15$ pJ m$^{-1}$, $D=3$ mJ m$^{-2}$, (effective) $K=0.425$ MJ m$^{-3}$, and $\alpha=0.1$, so that $l_{ex}=8.42$ nm, $t_0=7.8$ ps, $\xi=1.68$, and $\kappa=2.0$. Numerically, a $(40\times 40)\;l_{ex}^2$ calculation window with periodic boundary conditions and mesh cells of $(0.5\times 0.5)\;l_{ex}^2$ has been used.

\begin{figure}[!t]
	\centering
	\includegraphics[width=0.46\textwidth]{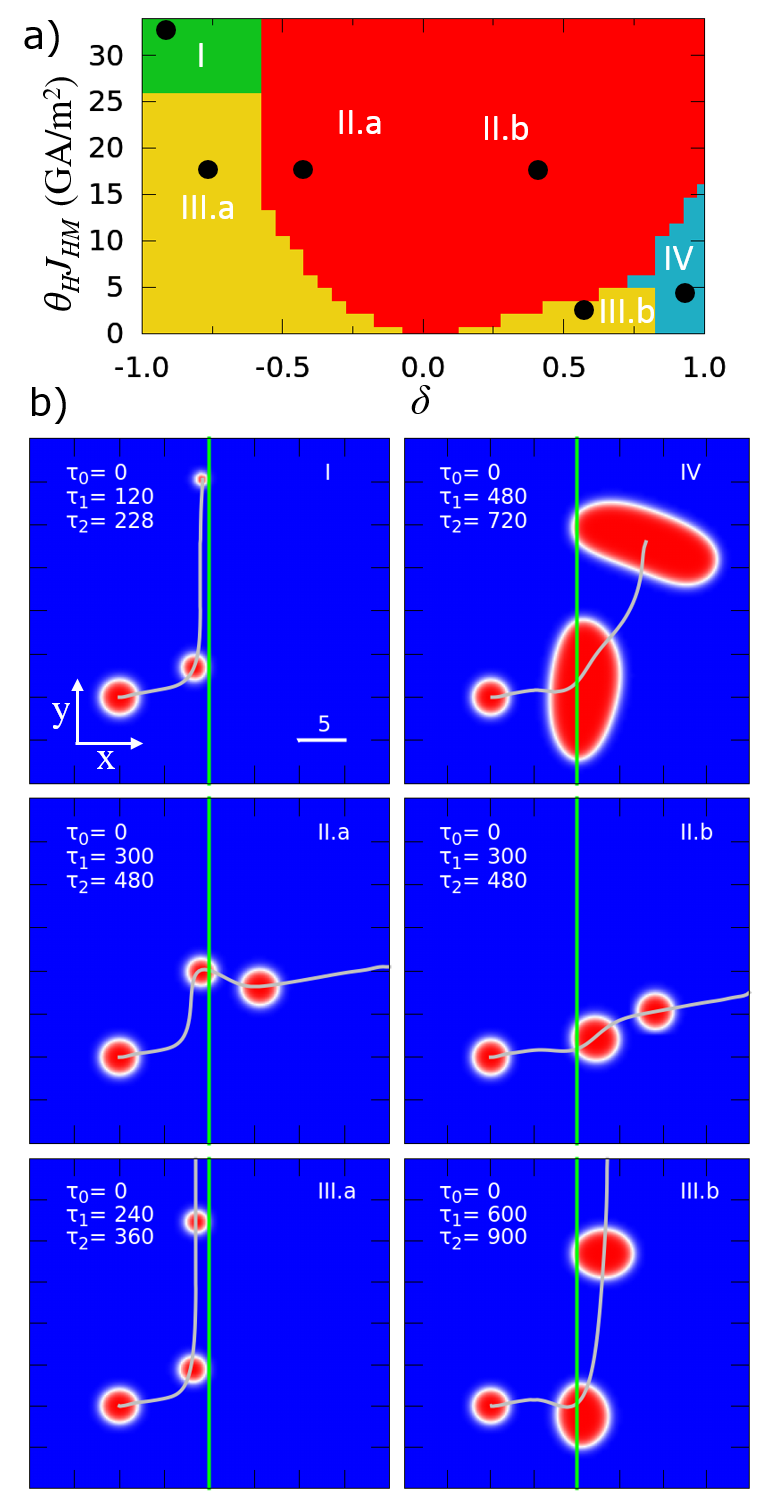}
	\caption{(a)$(\delta,\theta_H J_{HM})$-phase diagram for the different regimes of the skyrmion with a line defect (see text). (b) Snapshots at different times (indicated in each figure in normalized units, the leftmost structure corresponds to $\tau_0$) of the skyrmionic structure [$z$-component of the magnetization: blue (red) color corresponds to magnetization pointing $-{\bf \hat{z}}$ ($+{\bf \hat{z}}$), white to $m_z=0$]. Each plot is calculated with the parameters indicated by dots in the phase diagram. The green vertical line represents the defect and the gray line the trajectory of the skyrmion. The length scale is identical in all plots and is indicated in units of $l_{ex}$ in the top-left plot. Videos are available in the Electronic Supplementary Information.
	}
	\label{fig:PhaseDiagram}
\end{figure}

In the simulations we proceed in the following way. Initially, a uniform magnetization state in the $-{\bf \hat{z}}$ direction is considered. A skyrmion is generated by causing a distortion in the uniform background in the absence of current. The system evolves [Eq. \eqref{eq.LLG}] up to a given stationary skyrmionic state. With the parameters used the radius of the obtained skyrmion is $R\simeq 2.5\,l_{ex}\simeq 21$ nm. Then, the effective electronic current is considered to flow through the HM in the $y$-direction $\theta_H {\bf J}_{HM}= \theta_H J_{HM} {\bf \hat{y}}$ (where $\theta_H 
J_{HM}>0)$, so ${\bf \hat{\sigma}}={\bf \hat{x}}$. The skyrmion is driven by this current and moves towards the line defect.

\section{Results for a single line defect}
We first consider a skyrmion approaching a single line defect, which is regarded as a region with modified iDM interaction (with respect to the rest of the sample), in a similar way as in \cite{Navau2018,Menezes2019,Toscano2019}. Although modifying locally the iDM interaction could induce changes in other interactions, here we focus only on variations of iDM to avoid introducing extra parameters in the model \footnote{We have performed simulations considering local variations of anisotropy in addition to the iDM interation. The results found indicate that for variation factors similar to those used for iDM, there are no significant changes in the main conclusions.}. We assume that these defects are much more intense than other randomly distributed defects that could be present in the material. We are thus assuming that the skyrmions move in the flow regime, where their speed $v_0$ is proportional to $J_{HM}$ in the absence of line defects, and their direction has an angle with respect to the direction of ${\bf J}_{HM}$ (skyrmionic Hall effect) \cite{Tomasello2014,Jiang2016}. The presence of granularity or randomly distributed defects would set up in practice a lower bound for the current to be applied in order to move the skyrmion towards the defect  \cite{Kim2017APL,Raposo2017,Legrand2017}. This bound depends on the size of skyrmions and material parameters.  We are interested here in the acceleration properties of the created line defects. Thus, we assume that we are above this bound and focus on the effect that the purposely created line defects produce over the skyrmionic speed with respect to flowing speed in a plain defect-free sample.

Micromagnetically, the value  $\xi$  is changed in the defect positions into $\xi_d= \xi(1+\delta)$, where $\delta$ indicates the strength of the defect. The values of $\delta$ range from $\delta=-1$ (total suppression of the local iDM interaction on the line) to $\delta=1$ (doubling the local iDM interaction). $\delta<0$ ($\delta>0$) characterizes repulsive (attractive) defects \cite{Navau2018}. The kind of defects necessary for creating these rails could be obtained by locally modifying the spin-orbit coupling with different materials \cite{Moreau-Luchaire2016,Boulle2016,Ma2016,Hsu2018}, voltage gating using Brillouin light spectroscopy \cite{Srivastava2018}, inducing interfacial defects \cite{Romming2013}, current injection \cite{LimaFernandes2018}, lithographic techniques \cite{Balk2017, Wells2017}. Here we are considering defects at the atomic scale (as those in Refs. \cite{LimaFernandes2018,Stosic2017}, for example) and perfectly aligned. Considering that the radius of the skyrmion is much larger than the interatomic space, one expects that few-atom deviations from the perfect straight line and/or defects acting over few atomic positions would not affect significantly the validity of our model.

We show in Fig. \ref{fig:PhaseDiagram}a the calculated $(\delta,\theta_H J_{HM}$)-phase diagram showing different regimes. The results indicate that depending on the effective current density $\theta_H J_{HM}$ and  the strength of the defect $\delta$, the skyrmion can be trapped, annihilated (by implosion), destabilized (by continuous deformation), or else traverses the line defect. 

For weak defects ($\delta\sim 0$) the skyrmion goes through the defect. This region of crossing (regimes II.a and II.b in Fig. \ref{fig:PhaseDiagram}a) extends to larger $\delta$'s when increasing  $\theta_H J_{HM}$'s.  When the current density is large enough and/or when the defect is too strong (regions I and IV in Fig. \ref{fig:PhaseDiagram}a), other possibilities appear. In particular, the skyrmion can vanish when approaching the repulsive line defect (Fig. \ref{fig:PhaseDiagram}b-I) or can be largely distorted and eventually destroyed when passing through a strong attractive defect (Fig. \ref{fig:PhaseDiagram}b-IV). These cases are not useful for guiding skyrmions and will not be studied in detail here. Nevertheless, the presence of these regions sets up the limiting parameters for an effective guiding of skyrmions without destroying them. Note that as a general fact, when the skyrmion approaches the attractive (repulsive) defect its radius increases (decreases).

There are regions in the $(\delta,\theta_H J_{HM})$-phase diagram (III.a and III.b in Fig. \ref{fig:PhaseDiagram}a) where the skyrmion becomes trapped in the $x$ direction, but can move along the $y$ direction of the line defect.  Interestingly, the skyrmion can be trapped both if the defect is repulsive (Fig. \ref{fig:PhaseDiagram}b-III.a, $\delta<0$) or attractive (Fig. \ref{fig:PhaseDiagram}b-III.b, $\delta>0$). However, an important difference appears between the two cases. For the attractive defect, the skyrmion traverses the defect before being guided along it, in the $+{\bf \hat{y}}$ direction. If the defect is repulsive, the skyrmion does not cross the line defect before being guided in the \textit{same} direction  $+{\bf \hat{y}}$. Actually, if the skyrmion crossed a repulsive defect, it would escape. As a result, within the adequate range of driving velocities, both opposite defects push the skyrmion along the same direction. The other important effect is that the speed of the skyrmion, when trapped and guided along the defect, increases with respect to the initial skyrmion speed. This is true for both repulsive and attractive defects. 

To understand this behavior, consider the different forces acting over the skyrmion. For a rigid skyrmion, its velocity  ${\bf v}$ can be found using the Thiele's equation \cite{Ivanov2007,Liu2007}. Already simplified for our particular system, it reads
\begin{equation}
-G \hat{\bf z} \times {\bf v} - \alpha D {\bf v} - B \theta_H J_{HM} \hat{{\bf y}} + f(x) \hat{\bf x} = 0,
\label{eq.thiele}
\end{equation}
where the $f(x)$ function represents the forces created by the defect line \cite{Navau2018}, and $G$, $D$, and $B$ are positive constants that depend on the particular shape of the skyrmion. Due to the Magnus term [$G$-term in Eq. \eqref{eq.thiele}], both the current term [$J_{HM}$-term in Eq. \eqref{eq.thiele}] and the defect term [$f(x)$-term in Eq. \eqref{eq.thiele}] yield velocity components in both $x$ and $y$ directions. Eq. \eqref{eq.thiele} can be formally solved for the velocity components as
\begin{eqnarray}
v_x = \frac{1}{\alpha ^2 D^2+G^2}\left(B G \theta_H J_{HM} +\alpha  D f(x)\right), \\
v_y = \frac{1}{\alpha ^2 D^2+G^2}\left(\alpha  B D \theta_H J_{HM} - G f(x)\right).
\end{eqnarray}
For a repulsive line defect located at $x=0$, $f(x)<0$ for $x<0$ and $f(x)>0$ for $x>0$. This means that, if the skyrmion does not cross the line defect, $v_x$ can be canceled, so that the skyrmion is trapped in $x$; at the same time, $v_y$ contains additive contributions from the current and the defect terms. If the line defect is attractive, $f(x)>0$ for $x<0$ and $f(x)<0$ for $x>0$,  the compensation of $v_x$ can be done only after crossing the defect. In this case, both terms contribute again additively to $v_y$. 

The above rigid model does not account for the change in size of skyrmions when approaching to a line defect and this effect can also affect the velocity \cite{Legrand2017} (larger skyrmions move at larger velocities). From our micromagnetic calculations, the size of the skyrmion when guided along repulsive defects decreases, but its speed is remarkably increased, showing that the effect of the line defect over the velocity is the dominant effect. When the line defect is attractive, the skyrmion size increases slightly, which may even help in the speed-up.
 
All these results indicate that an optimum channel for skyrmions would be a combination of attractive-repulsive lines of defects.
It is important that the skyrmion first reaches the attractive defect, goes through it and then finds the repulsive defect and remains confined between both. In this arrangement, the skyrmion moves along the lane determined by the two rails and both cooperate in increasing the speed of the skyrmion. This is in contrast with the skyrmion moving on a track. There, both borders are repulsive and the skyrmion is accelerated when approaching one border but decelerated when approaching the other one \cite{Navau2016,Iwasaki2014}. 

\section{Results for a two line defects}
We now study the behavior of skyrmions when guided along rails formed by two line defects.
The attractive and repulsive line defects are characterized by $\xi_+$ and $\xi_-$, respectively. To simplify the treatment and reduce the number of parameters we characterize both line defects by a single parameter: $\xi_\pm=\xi (1\pm \delta)$. For the present study we have fixed $\theta_H J_{HM}=10$ GA m$^{-2}$, which gives a speed of skyrmions when no defects are considered of $v_0=22$ m/s.

\begin{figure}[t]
	\centering
		\includegraphics[width=0.55\textwidth]{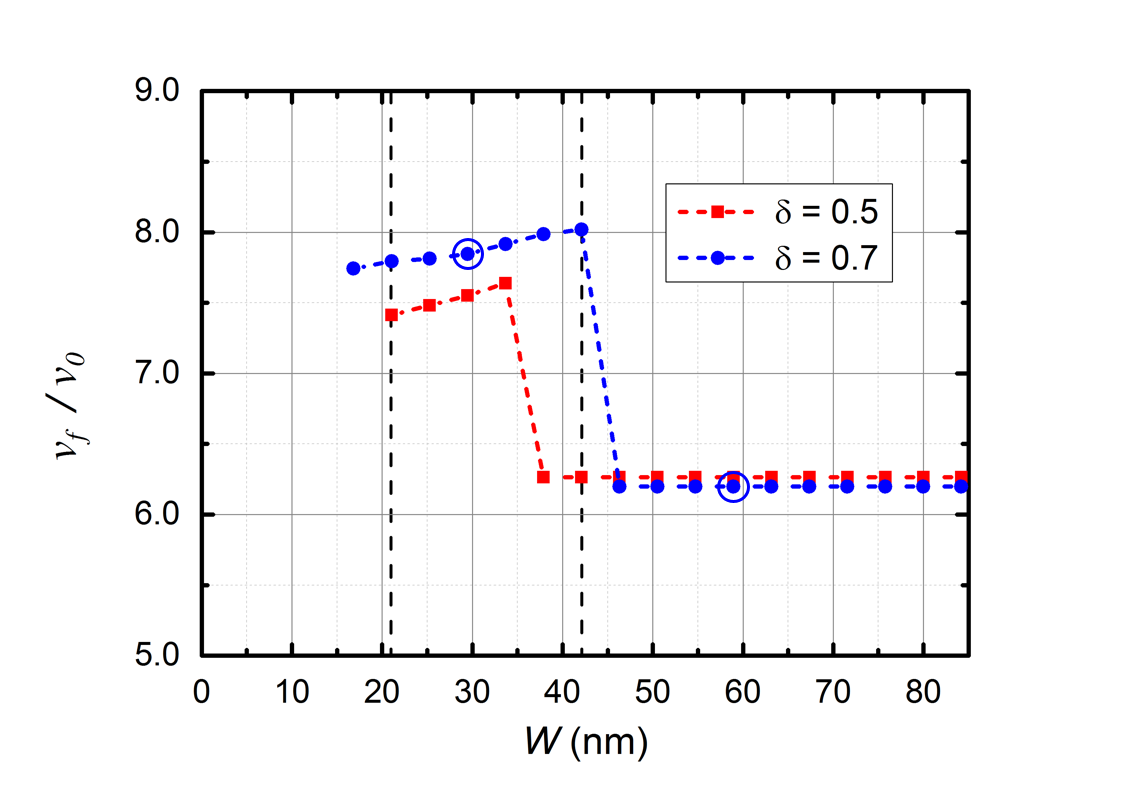}
	\caption{Ratio of the speed of skyrmions along the rails $v_f$ with respect to its initial speed $v_0$, as a function of the separation $W$ between the rails. The different lines correspond to different values of $\delta$. The vertical lines are shown for comparison and indicate the radius and diameter of the initial (far from the defects) skyrmion. The encircled points are the ones used for the snapshots in Fig. \ref{fig:2LD}.}
	\label{fig:optim}
\end{figure}

\begin{figure}[t]
	\centering
		\includegraphics[width=0.45\textwidth]{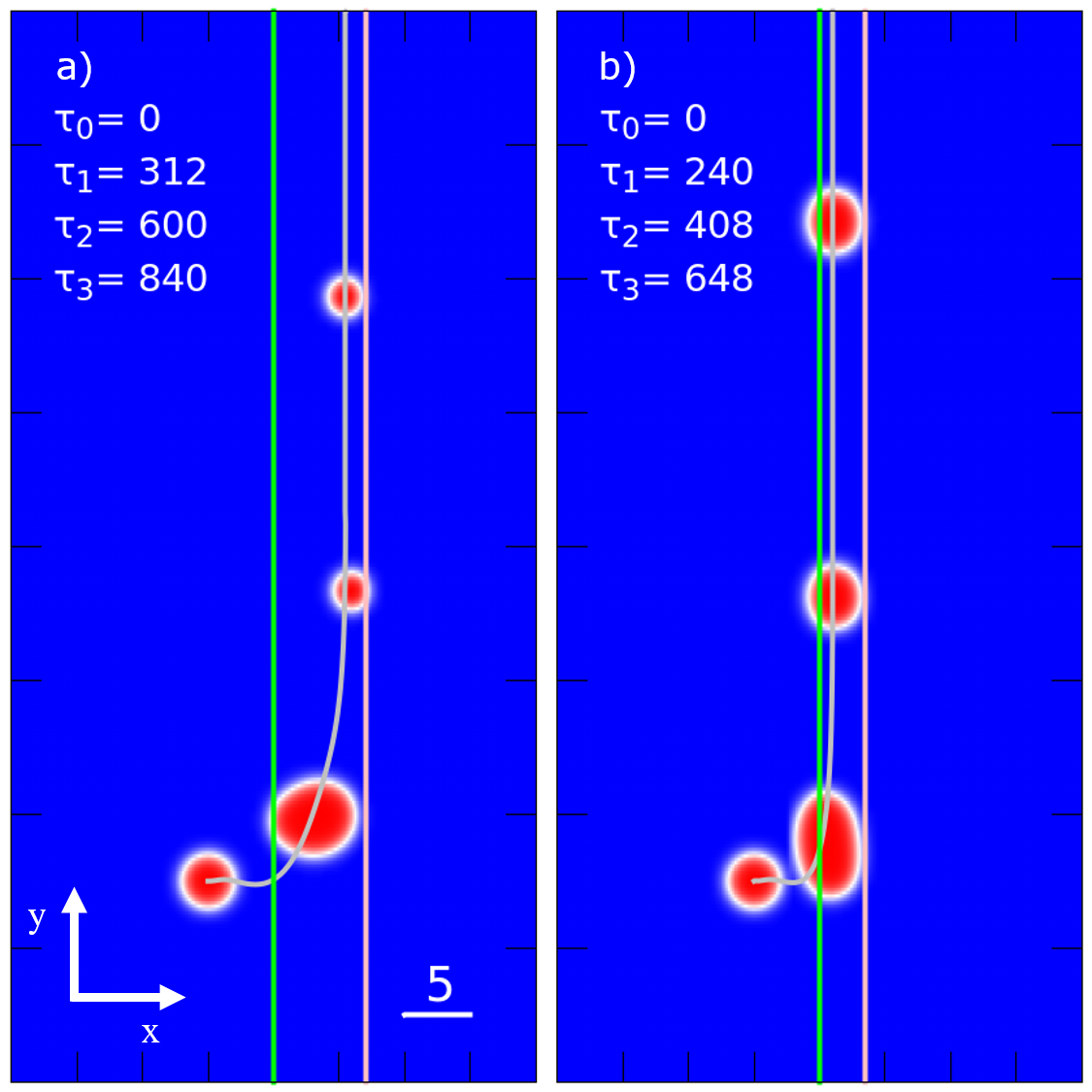}
	\caption{Snapshots of the skyrmion channeled between two line defects at different times indicated in the figure (in normalized units, the leftmost structure corresponds to $\tau_0$). The length scale is indicated in units of $l_{ex}$. Green (pink) vertical lines indicates the attractive (repulsive) line defect. The gray line is the trajectory of the skyrmion. The two plots correspond to different widths of the rails $W$. Note that the time interval between the two last snapshots is the same in both plots, showing the different speed-up factor for different widths of the channel. The parameters used are those corresponding to the encircled points in Fig. \ref{fig:optim}. Videos are available in the Electronic Supplementary Information.}
	\label{fig:2LD}
\end{figure}

The speed-up factor of the velocity along the channel with respect $v_0$, $v_f/v_0$, as a function of the separation between the two rails $W$ is shown in Fig. \ref{fig:optim} for the cases $\delta=0.7$ and 0.5. When $W$ is larger than the initial (far from the defects) diameter of the skyrmion, the skyrmion is guided due to the repulsive line defect with an speed-up factor of $\simeq 6$, see Fig. \ref{fig:2LD}a.  When  $W$ is reduced to approximately the diameter of the skyrmion an abrupt increase in the speed-up factor  is produced. The skyrmion is still trapped due to the repulsive defect. However, the additional contribution of the attractive line defect results in an increase of the skyrmion size and of the speed-up factor (the latter up to a value $\simeq 8$), see Fig. \ref{fig:2LD}b. For smaller $W$'s, the velocity is basically maintained up to a lower critical width $W_c$ of the order of the skyrmion radius (a small reduction in $v_f/v_0$ is seen mainly due to the different deformation that the skyrmion suffers). When $W<W_c$ the skyrmion escapes from the line defects since the two defects tend to cancel each other. By rising $\delta$, the jump in the $v_f/v_0$ factor moves to larger $W$ and, at the same time $W_c$ moves to lower values, thus increasing the observed region of higher $v_f/v_0$. Note however, that the value of $\delta$ cannot grow unlimitedly, since skyrmions would be no longer guided but destroyed by the defects themselves. For the case in Fig. \ref{fig:2LD}b, we have also performed calculations using different $\alpha$ values. We have found a $1/\alpha$ dependence of the speed-up factor. This result was also found in Ref. \cite{Iwasaki2014} for guiding along a repulsive borders and agrees with the Thiele's equation predictions.

The speed-up factor could be increased by reducing the effective applied current $\theta_H J_{HM}$, although this could result in creep effects before reaching the defect lines \cite{Kim2017APL,Raposo2017,Legrand2017}. Also, by considering different $\delta$'s for the two line defects one could optimize them to increase the speed-up factor. Actually, the obtained $v_f$ is around 175 m/s with $\theta_H J_{HM}=10$ GA/m$^2$. Although a direct comparison with other published data is difficult due to the variety of parameters used in the literature, we note that with the same $\theta_H J_{HM}$ and similar material parameters our $v_f$ is about 4 times larger than in \cite{He2017}. An alternative result is that with the use of rails, a given velocity for skyrmions can be achieved with a factor-$v_f/v_0$ less current density than without using them. In \cite{Chen2017,Song2017} approximately the same velocity for skyrmions was obtained using similar material parameters but requiring a current density an order of magnitude larger than that used  here. A further advantage of our proposal is that, in contrast to conventional scenarios \cite{Tomasello2014}, ${\bf J}_{HM}$ is parallel to the guiding rails, avoiding the skyrmionic Hall effect. This can be beneficial as one could reduce the total current needed, thus enhancing efficiency.

\section{Conclusions}
We have presented a novel platform for accelerating skyrmions along guiding rails. The rails consist of a pair of line defects, one of them attractive (local increase of iDM interaction) and the other repulsive (local reduction of iDM interaction). Within a range of driving current densities and strength of the defects the skyrmion is guided along channels with a width of the order of the radius of the initial skyrmion, whereas its speed is enhanced by an order of magnitude.  The use of skyrmions as information carriers will need the feasible control of writing, transporting, storing, and reading of individual skyrmions in a fast, dense, and controllable way. The present work provides a new strategy for enhancing the transport stage. One can also envisage parallelization of rails by tuning adequately the strength of the defects to achieve the desired velocity and size. The deformations would not compromise the carried information because the topological structure would not be destroyed. Being guided, accelerated and compressed, the amount of skyrmionic information that can be transported for a given total width of the film could be boosted. 

\section*{Acknowledgements}
We thank financial support from Catalan project 2017-SGR-105, 
Spanish project MAT2016-79426-P of Agencia Estatal de Investigaci\'on / Fondo Europeo de Desarrollo Regional (UE). A. S. acknowledges a grant from ICREA Academia, funded by the Generalitat de Catalunya. J. C.-Q. acknowledges a grant (FPU17/01970) from Ministerio de Ciencia, Innovaci\'on y Universidades (Spanish Government). 
 
\bibliography{channeling,bibothers}

\begin{thebibliography}{49}%
\makeatletter
\providecommand \@ifxundefined [1]{%
 \@ifx{#1\undefined}
}%
\providecommand \@ifnum [1]{%
 \ifnum #1\expandafter \@firstoftwo
 \else \expandafter \@secondoftwo
 \fi
}%
\providecommand \@ifx [1]{%
 \ifx #1\expandafter \@firstoftwo
 \else \expandafter \@secondoftwo
 \fi
}%
\providecommand \natexlab [1]{#1}%
\providecommand \enquote  [1]{``#1''}%
\providecommand \bibnamefont  [1]{#1}%
\providecommand \bibfnamefont [1]{#1}%
\providecommand \citenamefont [1]{#1}%
\providecommand \href@noop [0]{\@secondoftwo}%
\providecommand \href [0]{\begingroup \@sanitize@url \@href}%
\providecommand \@href[1]{\@@startlink{#1}\@@href}%
\providecommand \@@href[1]{\endgroup#1\@@endlink}%
\providecommand \@sanitize@url [0]{\catcode `\\12\catcode `\$12\catcode
  `\&12\catcode `\#12\catcode `\^12\catcode `\_12\catcode `\%12\relax}%
\providecommand \@@startlink[1]{}%
\providecommand \@@endlink[0]{}%
\providecommand \url  [0]{\begingroup\@sanitize@url \@url }%
\providecommand \@url [1]{\endgroup\@href {#1}{\urlprefix }}%
\providecommand \urlprefix  [0]{URL }%
\providecommand \Eprint [0]{\href }%
\providecommand \doibase [0]{http://dx.doi.org/}%
\providecommand \selectlanguage [0]{\@gobble}%
\providecommand \bibinfo  [0]{\@secondoftwo}%
\providecommand \bibfield  [0]{\@secondoftwo}%
\providecommand \translation [1]{[#1]}%
\providecommand \BibitemOpen [0]{}%
\providecommand \bibitemStop [0]{}%
\providecommand \bibitemNoStop [0]{.\EOS\space}%
\providecommand \EOS [0]{\spacefactor3000\relax}%
\providecommand \BibitemShut  [1]{\csname bibitem#1\endcsname}%
\let\auto@bib@innerbib\@empty
%</preamble>
\bibitem [{\citenamefont {Fert}\ \emph {et~al.}(2013)\citenamefont {Fert},
  \citenamefont {Cros},\ and\ \citenamefont {Sampaio}}]{Fert2013}%
  \BibitemOpen
  \bibfield  {author} {\bibinfo {author} {\bibfnamefont {A.}~\bibnamefont
  {Fert}}, \bibinfo {author} {\bibfnamefont {V.}~\bibnamefont {Cros}}, \ and\
  \bibinfo {author} {\bibfnamefont {J.}~\bibnamefont {Sampaio}},\ }\href
  {\doibase 10.1038/nnano.2013.29} {\bibfield  {journal} {\bibinfo  {journal}
  {Nature Nanotechnology}\ }\textbf {\bibinfo {volume} {8}},\ \bibinfo {pages}
  {152} (\bibinfo {year} {2013})}\BibitemShut {NoStop}%
\bibitem [{\citenamefont {Sampaio}\ \emph {et~al.}(2013)\citenamefont
  {Sampaio}, \citenamefont {Cros}, \citenamefont {Rohart}, \citenamefont
  {Thiaville},\ and\ \citenamefont {Fert}}]{Sampaio2013}%
  \BibitemOpen
  \bibfield  {author} {\bibinfo {author} {\bibfnamefont {J.}~\bibnamefont
  {Sampaio}}, \bibinfo {author} {\bibfnamefont {V.}~\bibnamefont {Cros}},
  \bibinfo {author} {\bibfnamefont {S.}~\bibnamefont {Rohart}}, \bibinfo
  {author} {\bibfnamefont {A.}~\bibnamefont {Thiaville}}, \ and\ \bibinfo
  {author} {\bibfnamefont {A.}~\bibnamefont {Fert}},\ }\href {\doibase
  10.1038/nnano.2013.210} {\bibfield  {journal} {\bibinfo  {journal} {Nature
  Nanotechnology}\ }\textbf {\bibinfo {volume} {8}},\ \bibinfo {pages} {839}
  (\bibinfo {year} {2013})}\BibitemShut {NoStop}%
\bibitem [{\citenamefont {Tomasello}\ \emph {et~al.}(2014)\citenamefont
  {Tomasello}, \citenamefont {Martinez}, \citenamefont {Zivieri}, \citenamefont
  {Torres}, \citenamefont {Carpentieri},\ and\ \citenamefont
  {Finocchio}}]{Tomasello2014}%
  \BibitemOpen
  \bibfield  {author} {\bibinfo {author} {\bibfnamefont {R.}~\bibnamefont
  {Tomasello}}, \bibinfo {author} {\bibfnamefont {E.}~\bibnamefont {Martinez}},
  \bibinfo {author} {\bibfnamefont {R.}~\bibnamefont {Zivieri}}, \bibinfo
  {author} {\bibfnamefont {L.}~\bibnamefont {Torres}}, \bibinfo {author}
  {\bibfnamefont {M.}~\bibnamefont {Carpentieri}}, \ and\ \bibinfo {author}
  {\bibfnamefont {G.}~\bibnamefont {Finocchio}},\ }\href {\doibase
  10.1038/srep06784} {\bibfield  {journal} {\bibinfo  {journal} {Scientific
  reports}\ }\textbf {\bibinfo {volume} {4}},\ \bibinfo {pages} {6784}
  (\bibinfo {year} {2014})}\BibitemShut {NoStop}%
\bibitem [{\citenamefont {Iwasaki}\ \emph {et~al.}(2013)\citenamefont
  {Iwasaki}, \citenamefont {Mochizuki},\ and\ \citenamefont
  {Nagaosa}}]{Iwasaki2013NatNano}%
  \BibitemOpen
  \bibfield  {author} {\bibinfo {author} {\bibfnamefont {J.}~\bibnamefont
  {Iwasaki}}, \bibinfo {author} {\bibfnamefont {M.}~\bibnamefont {Mochizuki}},
  \ and\ \bibinfo {author} {\bibfnamefont {N.}~\bibnamefont {Nagaosa}},\ }\href
  {\doibase 10.1038/nnano.2013.176} {\bibfield  {journal} {\bibinfo  {journal}
  {Nature Nanotechnology}\ }\textbf {\bibinfo {volume} {8}},\ \bibinfo {pages}
  {742} (\bibinfo {year} {2013})}\BibitemShut {NoStop}%
\bibitem [{\citenamefont {Woo}\ \emph {et~al.}(2016)\citenamefont {Woo},
  \citenamefont {Litzius}, \citenamefont {Kr{\"{u}}ger}, \citenamefont {Im},
  \citenamefont {Caretta}, \citenamefont {Richter}, \citenamefont {Mann},
  \citenamefont {Krone}, \citenamefont {Reeve}, \citenamefont {Weigand},
  \citenamefont {Agrawal}, \citenamefont {Lemesh}, \citenamefont {Mawass},
  \citenamefont {Fischer}, \citenamefont {Kl{\"{a}}ui},\ and\ \citenamefont
  {Beach}}]{Woo2016}%
  \BibitemOpen
  \bibfield  {author} {\bibinfo {author} {\bibfnamefont {S.}~\bibnamefont
  {Woo}}, \bibinfo {author} {\bibfnamefont {K.}~\bibnamefont {Litzius}},
  \bibinfo {author} {\bibfnamefont {B.}~\bibnamefont {Kr{\"{u}}ger}}, \bibinfo
  {author} {\bibfnamefont {M.-Y.}\ \bibnamefont {Im}}, \bibinfo {author}
  {\bibfnamefont {L.}~\bibnamefont {Caretta}}, \bibinfo {author} {\bibfnamefont
  {K.}~\bibnamefont {Richter}}, \bibinfo {author} {\bibfnamefont
  {M.}~\bibnamefont {Mann}}, \bibinfo {author} {\bibfnamefont {A.}~\bibnamefont
  {Krone}}, \bibinfo {author} {\bibfnamefont {R.~M.}\ \bibnamefont {Reeve}},
  \bibinfo {author} {\bibfnamefont {M.}~\bibnamefont {Weigand}}, \bibinfo
  {author} {\bibfnamefont {P.}~\bibnamefont {Agrawal}}, \bibinfo {author}
  {\bibfnamefont {I.}~\bibnamefont {Lemesh}}, \bibinfo {author} {\bibfnamefont
  {M.-A.}\ \bibnamefont {Mawass}}, \bibinfo {author} {\bibfnamefont
  {P.}~\bibnamefont {Fischer}}, \bibinfo {author} {\bibfnamefont
  {M.}~\bibnamefont {Kl{\"{a}}ui}}, \ and\ \bibinfo {author} {\bibfnamefont
  {G.~S.~D.}\ \bibnamefont {Beach}},\ }\href {\doibase 10.1038/nmat4593}
  {\bibfield  {journal} {\bibinfo  {journal} {Nature Materials}\ }\textbf
  {\bibinfo {volume} {15}},\ \bibinfo {pages} {501} (\bibinfo {year}
  {2016})}\BibitemShut {NoStop}%
\bibitem [{\citenamefont {Navau}\ \emph {et~al.}(2016)\citenamefont {Navau},
  \citenamefont {Del-Valle},\ and\ \citenamefont {Sanchez}}]{Navau2016}%
  \BibitemOpen
  \bibfield  {author} {\bibinfo {author} {\bibfnamefont {C.}~\bibnamefont
  {Navau}}, \bibinfo {author} {\bibfnamefont {N.}~\bibnamefont {Del-Valle}}, \
  and\ \bibinfo {author} {\bibfnamefont {A.}~\bibnamefont {Sanchez}},\ }\href
  {\doibase 10.1103/PhysRevB.94.184104} {\bibfield  {journal} {\bibinfo
  {journal} {Physical Review B}\ }\textbf {\bibinfo {volume} {94}},\ \bibinfo
  {pages} {184104} (\bibinfo {year} {2016})}\BibitemShut {NoStop}%
\bibitem [{\citenamefont {Chen}\ \emph {et~al.}(2017)\citenamefont {Chen},
  \citenamefont {Kang}, \citenamefont {Zhu}, \citenamefont {Zhang},
  \citenamefont {Lei}, \citenamefont {Zhang}, \citenamefont {Zhou},\ and\
  \citenamefont {Zhao}}]{Chen2017}%
  \BibitemOpen
  \bibfield  {author} {\bibinfo {author} {\bibfnamefont {X.}~\bibnamefont
  {Chen}}, \bibinfo {author} {\bibfnamefont {W.}~\bibnamefont {Kang}}, \bibinfo
  {author} {\bibfnamefont {D.}~\bibnamefont {Zhu}}, \bibinfo {author}
  {\bibfnamefont {X.}~\bibnamefont {Zhang}}, \bibinfo {author} {\bibfnamefont
  {N.}~\bibnamefont {Lei}}, \bibinfo {author} {\bibfnamefont {Y.}~\bibnamefont
  {Zhang}}, \bibinfo {author} {\bibfnamefont {Y.}~\bibnamefont {Zhou}}, \ and\
  \bibinfo {author} {\bibfnamefont {W.}~\bibnamefont {Zhao}},\ }\href {\doibase
  10.1063/1.5005953} {\bibfield  {journal} {\bibinfo  {journal} {Applied
  Physics Letters}\ }\textbf {\bibinfo {volume} {111}},\ \bibinfo {pages}
  {202406} (\bibinfo {year} {2017})}\BibitemShut {NoStop}%
\bibitem [{\citenamefont {Song}\ \emph {et~al.}(2017)\citenamefont {Song},
  \citenamefont {Jin}, \citenamefont {Wang}, \citenamefont {Xia}, \citenamefont
  {Wang},\ and\ \citenamefont {Liu}}]{Song2017}%
  \BibitemOpen
  \bibfield  {author} {\bibinfo {author} {\bibfnamefont {C.}~\bibnamefont
  {Song}}, \bibinfo {author} {\bibfnamefont {C.}~\bibnamefont {Jin}}, \bibinfo
  {author} {\bibfnamefont {J.}~\bibnamefont {Wang}}, \bibinfo {author}
  {\bibfnamefont {H.}~\bibnamefont {Xia}}, \bibinfo {author} {\bibfnamefont
  {J.}~\bibnamefont {Wang}}, \ and\ \bibinfo {author} {\bibfnamefont
  {Q.}~\bibnamefont {Liu}},\ }\href {\doibase 10.1063/1.4994093} {\bibfield
  {journal} {\bibinfo  {journal} {Applied Physics Letters}\ }\textbf {\bibinfo
  {volume} {111}},\ \bibinfo {pages} {192413} (\bibinfo {year}
  {2017})}\BibitemShut {NoStop}%
\bibitem [{\citenamefont {Zhang}\ \emph {et~al.}(2015)\citenamefont {Zhang},
  \citenamefont {Wang}, \citenamefont {Zheng}, \citenamefont {Zhu},
  \citenamefont {Liu}, \citenamefont {Chen}, \citenamefont {Jin}, \citenamefont
  {Liu}, \citenamefont {Jia},\ and\ \citenamefont {Xue}}]{Zhang2015NJP}%
  \BibitemOpen
  \bibfield  {author} {\bibinfo {author} {\bibfnamefont {S.}~\bibnamefont
  {Zhang}}, \bibinfo {author} {\bibfnamefont {J.}~\bibnamefont {Wang}},
  \bibinfo {author} {\bibfnamefont {Q.}~\bibnamefont {Zheng}}, \bibinfo
  {author} {\bibfnamefont {Q.}~\bibnamefont {Zhu}}, \bibinfo {author}
  {\bibfnamefont {X.}~\bibnamefont {Liu}}, \bibinfo {author} {\bibfnamefont
  {S.}~\bibnamefont {Chen}}, \bibinfo {author} {\bibfnamefont {C.}~\bibnamefont
  {Jin}}, \bibinfo {author} {\bibfnamefont {Q.}~\bibnamefont {Liu}}, \bibinfo
  {author} {\bibfnamefont {C.}~\bibnamefont {Jia}}, \ and\ \bibinfo {author}
  {\bibfnamefont {D.}~\bibnamefont {Xue}},\ }\href {\doibase
  10.1088/1367-2630/17/2/023061} {\bibfield  {journal} {\bibinfo  {journal}
  {New Journal of Physics}\ }\textbf {\bibinfo {volume} {17}},\ \bibinfo
  {pages} {023061} (\bibinfo {year} {2015})}\BibitemShut {NoStop}%
\bibitem [{\citenamefont {M{\"{u}}ller}(2017)}]{Muller2017}%
  \BibitemOpen
  \bibfield  {author} {\bibinfo {author} {\bibfnamefont {J.}~\bibnamefont
  {M{\"{u}}ller}},\ }\href {\doibase 10.1088/1367-2630/aa5b55} {\bibfield
  {journal} {\bibinfo  {journal} {New Journal of Physics}\ }\textbf {\bibinfo
  {volume} {19}},\ \bibinfo {pages} {025002} (\bibinfo {year}
  {2017})}\BibitemShut {NoStop}%
\bibitem [{\citenamefont {Purnama}\ \emph {et~al.}(2015)\citenamefont
  {Purnama}, \citenamefont {Gan}, \citenamefont {Wong},\ and\ \citenamefont
  {Lew}}]{Purnama2015}%
  \BibitemOpen
  \bibfield  {author} {\bibinfo {author} {\bibfnamefont {I.}~\bibnamefont
  {Purnama}}, \bibinfo {author} {\bibfnamefont {W.~L.}\ \bibnamefont {Gan}},
  \bibinfo {author} {\bibfnamefont {D.~W.}\ \bibnamefont {Wong}}, \ and\
  \bibinfo {author} {\bibfnamefont {W.~S.}\ \bibnamefont {Lew}},\ }\href
  {\doibase 10.1038/srep10620} {\bibfield  {journal} {\bibinfo  {journal}
  {Scientific Reports}\ }\textbf {\bibinfo {volume} {5}},\ \bibinfo {pages}
  {10620} (\bibinfo {year} {2015})}\BibitemShut {NoStop}%
\bibitem [{\citenamefont {Upadhyaya}\ \emph {et~al.}(2015)\citenamefont
  {Upadhyaya}, \citenamefont {Yu}, \citenamefont {Amiri},\ and\ \citenamefont
  {Wang}}]{Upadhyaya2015}%
  \BibitemOpen
  \bibfield  {author} {\bibinfo {author} {\bibfnamefont {P.}~\bibnamefont
  {Upadhyaya}}, \bibinfo {author} {\bibfnamefont {G.}~\bibnamefont {Yu}},
  \bibinfo {author} {\bibfnamefont {P.~K.}\ \bibnamefont {Amiri}}, \ and\
  \bibinfo {author} {\bibfnamefont {K.~L.}\ \bibnamefont {Wang}},\ }\href
  {\doibase 10.1103/PhysRevB.92.134411} {\bibfield  {journal} {\bibinfo
  {journal} {Physical Review B}\ }\textbf {\bibinfo {volume} {92}},\ \bibinfo
  {pages} {134411} (\bibinfo {year} {2015})}\BibitemShut {NoStop}%
\bibitem [{\citenamefont {Kang}\ \emph {et~al.}(2016)\citenamefont {Kang},
  \citenamefont {Huang}, \citenamefont {Zheng}, \citenamefont {Lv},
  \citenamefont {Lei}, \citenamefont {Zhang}, \citenamefont {Zhang},
  \citenamefont {Zhou},\ and\ \citenamefont {Zhao}}]{Kang2016}%
  \BibitemOpen
  \bibfield  {author} {\bibinfo {author} {\bibfnamefont {W.}~\bibnamefont
  {Kang}}, \bibinfo {author} {\bibfnamefont {Y.}~\bibnamefont {Huang}},
  \bibinfo {author} {\bibfnamefont {C.}~\bibnamefont {Zheng}}, \bibinfo
  {author} {\bibfnamefont {W.}~\bibnamefont {Lv}}, \bibinfo {author}
  {\bibfnamefont {N.}~\bibnamefont {Lei}}, \bibinfo {author} {\bibfnamefont
  {Y.}~\bibnamefont {Zhang}}, \bibinfo {author} {\bibfnamefont
  {X.}~\bibnamefont {Zhang}}, \bibinfo {author} {\bibfnamefont
  {Y.}~\bibnamefont {Zhou}}, \ and\ \bibinfo {author} {\bibfnamefont
  {W.}~\bibnamefont {Zhao}},\ }\href {\doibase 10.1038/srep23164} {\bibfield
  {journal} {\bibinfo  {journal} {Scientific Reports}\ }\textbf {\bibinfo
  {volume} {6}},\ \bibinfo {pages} {23164} (\bibinfo {year}
  {2016})}\BibitemShut {NoStop}%
\bibitem [{\citenamefont {Mulkers}\ \emph {et~al.}(2017)\citenamefont
  {Mulkers}, \citenamefont {{Van Waeyenberge}},\ and\ \citenamefont
  {Milo{\v{s}}evi{\'{c}}}}]{Mulkers2017}%
  \BibitemOpen
  \bibfield  {author} {\bibinfo {author} {\bibfnamefont {J.}~\bibnamefont
  {Mulkers}}, \bibinfo {author} {\bibfnamefont {B.}~\bibnamefont {{Van
  Waeyenberge}}}, \ and\ \bibinfo {author} {\bibfnamefont {M.~V.}\ \bibnamefont
  {Milo{\v{s}}evi{\'{c}}}},\ }\href {\doibase 10.1103/PhysRevB.95.144401}
  {\bibfield  {journal} {\bibinfo  {journal} {Physical Review B}\ }\textbf
  {\bibinfo {volume} {95}},\ \bibinfo {pages} {144401} (\bibinfo {year}
  {2017})}\BibitemShut {NoStop}%
\bibitem [{\citenamefont {Zhang}\ \emph {et~al.}(2017)\citenamefont {Zhang},
  \citenamefont {Luo}, \citenamefont {Yan}, \citenamefont {Ou-Yang},
  \citenamefont {Yang}, \citenamefont {Chen}, \citenamefont {Zhu},\ and\
  \citenamefont {You}}]{Zhang2017Nanoscale}%
  \BibitemOpen
  \bibfield  {author} {\bibinfo {author} {\bibfnamefont {Y.}~\bibnamefont
  {Zhang}}, \bibinfo {author} {\bibfnamefont {S.}~\bibnamefont {Luo}}, \bibinfo
  {author} {\bibfnamefont {B.}~\bibnamefont {Yan}}, \bibinfo {author}
  {\bibfnamefont {J.}~\bibnamefont {Ou-Yang}}, \bibinfo {author} {\bibfnamefont
  {X.}~\bibnamefont {Yang}}, \bibinfo {author} {\bibfnamefont {S.}~\bibnamefont
  {Chen}}, \bibinfo {author} {\bibfnamefont {B.}~\bibnamefont {Zhu}}, \ and\
  \bibinfo {author} {\bibfnamefont {L.}~\bibnamefont {You}},\ }\href {\doibase
  10.1039/C7NR01980G} {\bibfield  {journal} {\bibinfo  {journal} {Nanoscale}\
  }\textbf {\bibinfo {volume} {9}},\ \bibinfo {pages} {10212} (\bibinfo {year}
  {2017})}\BibitemShut {NoStop}%
\bibitem [{\citenamefont {Iwasaki}\ \emph {et~al.}(2014)\citenamefont
  {Iwasaki}, \citenamefont {Koshibae},\ and\ \citenamefont
  {Nagaosa}}]{Iwasaki2014}%
  \BibitemOpen
  \bibfield  {author} {\bibinfo {author} {\bibfnamefont {J.}~\bibnamefont
  {Iwasaki}}, \bibinfo {author} {\bibfnamefont {W.}~\bibnamefont {Koshibae}}, \
  and\ \bibinfo {author} {\bibfnamefont {N.}~\bibnamefont {Nagaosa}},\ }\href
  {\doibase 10.1021/nl501379k} {\bibfield  {journal} {\bibinfo  {journal} {Nano
  Letters}\ }\textbf {\bibinfo {volume} {14}},\ \bibinfo {pages} {4432}
  (\bibinfo {year} {2014})}\BibitemShut {NoStop}%
\bibitem [{\citenamefont {Stosic}\ \emph {et~al.}(2017)\citenamefont {Stosic},
  \citenamefont {Ludermir},\ and\ \citenamefont
  {Milo{\v{s}}evi{\'{c}}}}]{Stosic2017}%
  \BibitemOpen
  \bibfield  {author} {\bibinfo {author} {\bibfnamefont {D.}~\bibnamefont
  {Stosic}}, \bibinfo {author} {\bibfnamefont {T.~B.}\ \bibnamefont
  {Ludermir}}, \ and\ \bibinfo {author} {\bibfnamefont {M.~V.}\ \bibnamefont
  {Milo{\v{s}}evi{\'{c}}}},\ }\href {\doibase 10.1103/PhysRevB.96.214403}
  {\bibfield  {journal} {\bibinfo  {journal} {Physical Review B}\ }\textbf
  {\bibinfo {volume} {96}},\ \bibinfo {pages} {214403} (\bibinfo {year}
  {2017})}\BibitemShut {NoStop}%
\bibitem [{\citenamefont {Kim}\ \emph {et~al.}(2017)\citenamefont {Kim},
  \citenamefont {Lee},\ and\ \citenamefont {Tserkovnyak}}]{Kim2017PRB}%
  \BibitemOpen
  \bibfield  {author} {\bibinfo {author} {\bibfnamefont {S.~K.}\ \bibnamefont
  {Kim}}, \bibinfo {author} {\bibfnamefont {K.~J.}\ \bibnamefont {Lee}}, \ and\
  \bibinfo {author} {\bibfnamefont {Y.}~\bibnamefont {Tserkovnyak}},\ }\href
  {\doibase 10.1103/PhysRevB.95.140404} {\bibfield  {journal} {\bibinfo
  {journal} {Physical Review B}\ }\textbf {\bibinfo {volume} {95}},\ \bibinfo
  {pages} {140404} (\bibinfo {year} {2017})},\ \Eprint
  {http://arxiv.org/abs/1702.02554} {arXiv:1702.02554} \BibitemShut {NoStop}%
\bibitem [{\citenamefont {Navau}\ \emph {et~al.}(2018)\citenamefont {Navau},
  \citenamefont {Del-Valle},\ and\ \citenamefont {Sanchez}}]{Navau2018}%
  \BibitemOpen
  \bibfield  {author} {\bibinfo {author} {\bibfnamefont {C.}~\bibnamefont
  {Navau}}, \bibinfo {author} {\bibfnamefont {N.}~\bibnamefont {Del-Valle}}, \
  and\ \bibinfo {author} {\bibfnamefont {A.}~\bibnamefont {Sanchez}},\ }\href
  {\doibase 10.1016/J.JMMM.2018.06.044} {\bibfield  {journal} {\bibinfo
  {journal} {Journal of Magnetism and Magnetic Materials}\ }\textbf {\bibinfo
  {volume} {465}},\ \bibinfo {pages} {709} (\bibinfo {year}
  {2018})}\BibitemShut {NoStop}%
\bibitem [{\citenamefont {M{\"{u}}ller}\ and\ \citenamefont
  {Rosch}(2015)}]{Muller2015}%
  \BibitemOpen
  \bibfield  {author} {\bibinfo {author} {\bibfnamefont {J.}~\bibnamefont
  {M{\"{u}}ller}}\ and\ \bibinfo {author} {\bibfnamefont {A.}~\bibnamefont
  {Rosch}},\ }\href {\doibase 10.1103/PhysRevB.91.054410} {\bibfield  {journal}
  {\bibinfo  {journal} {Physical Review B}\ }\textbf {\bibinfo {volume} {91}},\
  \bibinfo {pages} {054410} (\bibinfo {year} {2015})}\BibitemShut {NoStop}%
\bibitem [{\citenamefont {Liu}\ and\ \citenamefont {Li}(2013)}]{Liu2013}%
  \BibitemOpen
  \bibfield  {author} {\bibinfo {author} {\bibfnamefont {Y.-H.}\ \bibnamefont
  {Liu}}\ and\ \bibinfo {author} {\bibfnamefont {Y.-Q.}\ \bibnamefont {Li}},\
  }\href {\doibase 10.1088/0953-8984/25/7/076005} {\bibfield  {journal}
  {\bibinfo  {journal} {Journal of Physics: Condensed Matter}\ }\textbf
  {\bibinfo {volume} {25}},\ \bibinfo {pages} {076005} (\bibinfo {year}
  {2013})}\BibitemShut {NoStop}%
\bibitem [{\citenamefont {{Lima Fernandes}}\ \emph {et~al.}(2018)\citenamefont
  {{Lima Fernandes}}, \citenamefont {Bouaziz}, \citenamefont {Bl{\"{u}}gel},\
  and\ \citenamefont {Lounis}}]{LimaFernandes2018}%
  \BibitemOpen
  \bibfield  {author} {\bibinfo {author} {\bibfnamefont {I.}~\bibnamefont
  {{Lima Fernandes}}}, \bibinfo {author} {\bibfnamefont {J.}~\bibnamefont
  {Bouaziz}}, \bibinfo {author} {\bibfnamefont {S.}~\bibnamefont
  {Bl{\"{u}}gel}}, \ and\ \bibinfo {author} {\bibfnamefont {S.}~\bibnamefont
  {Lounis}},\ }\href {\doibase 10.1038/s41467-018-06827-5} {\bibfield
  {journal} {\bibinfo  {journal} {Nature Communications}\ }\textbf {\bibinfo
  {volume} {9}},\ \bibinfo {pages} {4395} (\bibinfo {year} {2018})}\BibitemShut
  {NoStop}%
\bibitem [{\citenamefont {Liu}\ \emph {et~al.}(2011)\citenamefont {Liu},
  \citenamefont {Moriyama}, \citenamefont {Ralph},\ and\ \citenamefont
  {Buhrman}}]{Liu2011}%
  \BibitemOpen
  \bibfield  {author} {\bibinfo {author} {\bibfnamefont {L.}~\bibnamefont
  {Liu}}, \bibinfo {author} {\bibfnamefont {T.}~\bibnamefont {Moriyama}},
  \bibinfo {author} {\bibfnamefont {D.~C.}\ \bibnamefont {Ralph}}, \ and\
  \bibinfo {author} {\bibfnamefont {R.~A.}\ \bibnamefont {Buhrman}},\ }\href
  {\doibase 10.1103/PhysRevLett.106.036601} {\bibfield  {journal} {\bibinfo
  {journal} {Physical Review Letters}\ }\textbf {\bibinfo {volume} {106}},\
  \bibinfo {pages} {036601} (\bibinfo {year} {2011})}\BibitemShut {NoStop}%
\bibitem [{\citenamefont {Finocchio}\ \emph {et~al.}(2013)\citenamefont
  {Finocchio}, \citenamefont {Carpentieri}, \citenamefont {Martinez},\ and\
  \citenamefont {Azzerboni}}]{Finocchio2013}%
  \BibitemOpen
  \bibfield  {author} {\bibinfo {author} {\bibfnamefont {G.}~\bibnamefont
  {Finocchio}}, \bibinfo {author} {\bibfnamefont {M.}~\bibnamefont
  {Carpentieri}}, \bibinfo {author} {\bibfnamefont {E.}~\bibnamefont
  {Martinez}}, \ and\ \bibinfo {author} {\bibfnamefont {B.}~\bibnamefont
  {Azzerboni}},\ }\href {\doibase 10.1063/1.4808092} {\bibfield  {journal}
  {\bibinfo  {journal} {Applied Physics Letters}\ }\textbf {\bibinfo {volume}
  {102}},\ \bibinfo {pages} {212410} (\bibinfo {year} {2013})}\BibitemShut
  {NoStop}%
\bibitem [{\citenamefont {Knoester}\ \emph {et~al.}(2014)\citenamefont
  {Knoester}, \citenamefont {Sinova},\ and\ \citenamefont
  {Duine}}]{Knoester2014}%
  \BibitemOpen
  \bibfield  {author} {\bibinfo {author} {\bibfnamefont {M.~E.}\ \bibnamefont
  {Knoester}}, \bibinfo {author} {\bibfnamefont {J.}~\bibnamefont {Sinova}}, \
  and\ \bibinfo {author} {\bibfnamefont {R.~A.}\ \bibnamefont {Duine}},\ }\href
  {\doibase 10.1103/PhysRevB.89.064425} {\bibfield  {journal} {\bibinfo
  {journal} {Physical Review B}\ }\textbf {\bibinfo {volume} {89}},\ \bibinfo
  {pages} {064425} (\bibinfo {year} {2014})}\BibitemShut {NoStop}%
\bibitem [{\citenamefont {Zhang}\ and\ \citenamefont {Li}(2004)}]{Zhang2004}%
  \BibitemOpen
  \bibfield  {author} {\bibinfo {author} {\bibfnamefont {S.}~\bibnamefont
  {Zhang}}\ and\ \bibinfo {author} {\bibfnamefont {Z.}~\bibnamefont {Li}},\
  }\href {\doibase 10.1103/PhysRevLett.93.127204} {\bibfield  {journal}
  {\bibinfo  {journal} {Physical Review Letters}\ }\textbf {\bibinfo {volume}
  {93}},\ \bibinfo {pages} {127204} (\bibinfo {year} {2004})}\BibitemShut
  {NoStop}%
\bibitem [{\citenamefont {Coey}(2010)}]{Coey2010}%
  \BibitemOpen
  \bibfield  {author} {\bibinfo {author} {\bibfnamefont {J.~M.~D.}\
  \bibnamefont {Coey}},\ }\href {\doibase 10.1017/CBO9780511845000} {\emph
  {\bibinfo {title} {{Magnetism and Magnetic Materials}}}}\ (\bibinfo
  {publisher} {Cambridge University Press},\ \bibinfo {address} {Cambridge},\
  \bibinfo {year} {2010})\BibitemShut {NoStop}%
\bibitem [{\citenamefont {Raposo}\ \emph {et~al.}(2017)\citenamefont {Raposo},
  \citenamefont {{Luis Martinez}},\ and\ \citenamefont
  {Martinez}}]{Raposo2017}%
  \BibitemOpen
  \bibfield  {author} {\bibinfo {author} {\bibfnamefont {V.}~\bibnamefont
  {Raposo}}, \bibinfo {author} {\bibfnamefont {R.~F.}\ \bibnamefont {{Luis
  Martinez}}}, \ and\ \bibinfo {author} {\bibfnamefont {E.}~\bibnamefont
  {Martinez}},\ }\href {\doibase 10.1063/1.4975658} {\bibfield  {journal}
  {\bibinfo  {journal} {AIP Advances}\ }\textbf {\bibinfo {volume} {7}},\
  \bibinfo {pages} {056017} (\bibinfo {year} {2017})}\BibitemShut {NoStop}%
\bibitem [{\citenamefont {Metaxas}\ \emph {et~al.}(2007)\citenamefont
  {Metaxas}, \citenamefont {Jamet}, \citenamefont {Mougin}, \citenamefont
  {Cormier}, \citenamefont {Ferr{\'{e}}}, \citenamefont {Baltz}, \citenamefont
  {Rodmacq}, \citenamefont {Dieny},\ and\ \citenamefont
  {Stamps}}]{Metaxas2007}%
  \BibitemOpen
  \bibfield  {author} {\bibinfo {author} {\bibfnamefont {P.~J.}\ \bibnamefont
  {Metaxas}}, \bibinfo {author} {\bibfnamefont {J.~P.}\ \bibnamefont {Jamet}},
  \bibinfo {author} {\bibfnamefont {A.}~\bibnamefont {Mougin}}, \bibinfo
  {author} {\bibfnamefont {M.}~\bibnamefont {Cormier}}, \bibinfo {author}
  {\bibfnamefont {J.}~\bibnamefont {Ferr{\'{e}}}}, \bibinfo {author}
  {\bibfnamefont {V.}~\bibnamefont {Baltz}}, \bibinfo {author} {\bibfnamefont
  {B.}~\bibnamefont {Rodmacq}}, \bibinfo {author} {\bibfnamefont
  {B.}~\bibnamefont {Dieny}}, \ and\ \bibinfo {author} {\bibfnamefont {R.~L.}\
  \bibnamefont {Stamps}},\ }\href {\doibase 10.1103/PhysRevLett.99.217208}
  {\bibfield  {journal} {\bibinfo  {journal} {Physical Review Letters}\
  }\textbf {\bibinfo {volume} {99}},\ \bibinfo {pages} {217208} (\bibinfo
  {year} {2007})}\BibitemShut {NoStop}%
\bibitem [{\citenamefont {Del-Valle}\ \emph {et~al.}(2015)\citenamefont
  {Del-Valle}, \citenamefont {Agramunt-Puig}, \citenamefont {Sanchez},\ and\
  \citenamefont {Navau}}]{Del-Valle2015}%
  \BibitemOpen
  \bibfield  {author} {\bibinfo {author} {\bibfnamefont {N.}~\bibnamefont
  {Del-Valle}}, \bibinfo {author} {\bibfnamefont {S.}~\bibnamefont
  {Agramunt-Puig}}, \bibinfo {author} {\bibfnamefont {A.}~\bibnamefont
  {Sanchez}}, \ and\ \bibinfo {author} {\bibfnamefont {C.}~\bibnamefont
  {Navau}},\ }\href {\doibase 10.1063/1.4932090} {\bibfield  {journal}
  {\bibinfo  {journal} {Applied Physics Letters}\ }\textbf {\bibinfo {volume}
  {107}},\ \bibinfo {pages} {133103} (\bibinfo {year} {2015})}\BibitemShut
  {NoStop}%
\bibitem [{\citenamefont {Schellekens}\ \emph {et~al.}(2013)\citenamefont
  {Schellekens}, \citenamefont {Deen}, \citenamefont {Wang}, \citenamefont
  {Kohlhepp}, \citenamefont {Swagten},\ and\ \citenamefont
  {Koopmans}}]{Schellekens2013}%
  \BibitemOpen
  \bibfield  {author} {\bibinfo {author} {\bibfnamefont {A.~J.}\ \bibnamefont
  {Schellekens}}, \bibinfo {author} {\bibfnamefont {L.}~\bibnamefont {Deen}},
  \bibinfo {author} {\bibfnamefont {D.}~\bibnamefont {Wang}}, \bibinfo {author}
  {\bibfnamefont {J.~T.}\ \bibnamefont {Kohlhepp}}, \bibinfo {author}
  {\bibfnamefont {H.~J.~M.}\ \bibnamefont {Swagten}}, \ and\ \bibinfo {author}
  {\bibfnamefont {B.}~\bibnamefont {Koopmans}},\ }\href {\doibase
  10.1063/1.4794538} {\bibfield  {journal} {\bibinfo  {journal} {Applied
  Physics Letters}\ }\textbf {\bibinfo {volume} {102}},\ \bibinfo {pages}
  {082405} (\bibinfo {year} {2013})}\BibitemShut {NoStop}%
\bibitem [{\citenamefont {Belmeguenai}\ \emph {et~al.}(2015)\citenamefont
  {Belmeguenai}, \citenamefont {Adam}, \citenamefont {Roussign{\'{e}}},
  \citenamefont {Eimer}, \citenamefont {Devolder}, \citenamefont {Kim},
  \citenamefont {Cherif}, \citenamefont {Stashkevich},\ and\ \citenamefont
  {Thiaville}}]{Belmeguenai2015}%
  \BibitemOpen
  \bibfield  {author} {\bibinfo {author} {\bibfnamefont {M.}~\bibnamefont
  {Belmeguenai}}, \bibinfo {author} {\bibfnamefont {J.-P.}\ \bibnamefont
  {Adam}}, \bibinfo {author} {\bibfnamefont {Y.}~\bibnamefont
  {Roussign{\'{e}}}}, \bibinfo {author} {\bibfnamefont {S.}~\bibnamefont
  {Eimer}}, \bibinfo {author} {\bibfnamefont {T.}~\bibnamefont {Devolder}},
  \bibinfo {author} {\bibfnamefont {J.-V.}\ \bibnamefont {Kim}}, \bibinfo
  {author} {\bibfnamefont {S.~M.}\ \bibnamefont {Cherif}}, \bibinfo {author}
  {\bibfnamefont {A.}~\bibnamefont {Stashkevich}}, \ and\ \bibinfo {author}
  {\bibfnamefont {A.}~\bibnamefont {Thiaville}},\ }\href {\doibase
  10.1103/PhysRevB.91.180405} {\bibfield  {journal} {\bibinfo  {journal}
  {Physical Review B}\ }\textbf {\bibinfo {volume} {91}},\ \bibinfo {pages}
  {180405} (\bibinfo {year} {2015})}\BibitemShut {NoStop}%
\bibitem [{\citenamefont {Kim}\ and\ \citenamefont {Yoo}(2017)}]{Kim2017APL}%
  \BibitemOpen
  \bibfield  {author} {\bibinfo {author} {\bibfnamefont {J.-V.}\ \bibnamefont
  {Kim}}\ and\ \bibinfo {author} {\bibfnamefont {M.-W.}\ \bibnamefont {Yoo}},\
  }\href {\doibase 10.1063/1.4979316} {\bibfield  {journal} {\bibinfo
  {journal} {Applied Physics Letters}\ }\textbf {\bibinfo {volume} {110}},\
  \bibinfo {pages} {132404} (\bibinfo {year} {2017})}\BibitemShut {NoStop}%
\bibitem [{\citenamefont {Menezes}\ \emph {et~al.}(2019)\citenamefont
  {Menezes}, \citenamefont {Mulkers}, \citenamefont {Silva},\ and\
  \citenamefont {Milo{\v{s}}evi{\'{c}}}}]{Menezes2019}%
  \BibitemOpen
  \bibfield  {author} {\bibinfo {author} {\bibfnamefont {R.~M.}\ \bibnamefont
  {Menezes}}, \bibinfo {author} {\bibfnamefont {J.}~\bibnamefont {Mulkers}},
  \bibinfo {author} {\bibfnamefont {C.~C. d.~S.}\ \bibnamefont {Silva}}, \ and\
  \bibinfo {author} {\bibfnamefont {M.~V.}\ \bibnamefont
  {Milo{\v{s}}evi{\'{c}}}},\ }\href {\doibase 10.1103/PhysRevB.99.104409}
  {\bibfield  {journal} {\bibinfo  {journal} {Physical Review B}\ }\textbf
  {\bibinfo {volume} {99}},\ \bibinfo {pages} {104409} (\bibinfo {year}
  {2019})}\BibitemShut {NoStop}%
\bibitem [{\citenamefont {Toscano}\ \emph {et~al.}(2019)\citenamefont
  {Toscano}, \citenamefont {Leonel}, \citenamefont {Coura},\ and\ \citenamefont
  {Sato}}]{Toscano2019}%
  \BibitemOpen
  \bibfield  {author} {\bibinfo {author} {\bibfnamefont {D.}~\bibnamefont
  {Toscano}}, \bibinfo {author} {\bibfnamefont {S.}~\bibnamefont {Leonel}},
  \bibinfo {author} {\bibfnamefont {P.}~\bibnamefont {Coura}}, \ and\ \bibinfo
  {author} {\bibfnamefont {F.}~\bibnamefont {Sato}},\ }\href {\doibase
  10.1016/J.JMMM.2019.02.075} {\bibfield  {journal} {\bibinfo  {journal}
  {Journal of Magnetism and Magnetic Materials}\ }\textbf {\bibinfo {volume}
  {480}},\ \bibinfo {pages} {171} (\bibinfo {year} {2019})}\BibitemShut
  {NoStop}%
\bibitem [{Note1()}]{Note1}%
  \BibitemOpen
  \bibinfo {note} {We have performed simulations considering local variations
  of anisotropy in addition to the iDM interation. The results found indicate
  that for variation factors similar to those used for iDM, there are no
  significant changes in the main conclusions.}\BibitemShut {Stop}%
\bibitem [{\citenamefont {Jiang}\ \emph {et~al.}(2016)\citenamefont {Jiang},
  \citenamefont {Zhang}, \citenamefont {Yu}, \citenamefont {Zhang},
  \citenamefont {Wang}, \citenamefont {{Benjamin Jungfleisch}}, \citenamefont
  {Pearson}, \citenamefont {Cheng}, \citenamefont {Heinonen}, \citenamefont
  {Wang}, \citenamefont {Zhou}, \citenamefont {Hoffmann}, \citenamefont
  {te~Velthuis},\ and\ \citenamefont {E.}}]{Jiang2016}%
  \BibitemOpen
  \bibfield  {author} {\bibinfo {author} {\bibfnamefont {W.}~\bibnamefont
  {Jiang}}, \bibinfo {author} {\bibfnamefont {X.}~\bibnamefont {Zhang}},
  \bibinfo {author} {\bibfnamefont {G.}~\bibnamefont {Yu}}, \bibinfo {author}
  {\bibfnamefont {W.}~\bibnamefont {Zhang}}, \bibinfo {author} {\bibfnamefont
  {X.}~\bibnamefont {Wang}}, \bibinfo {author} {\bibfnamefont {M.}~\bibnamefont
  {{Benjamin Jungfleisch}}}, \bibinfo {author} {\bibfnamefont {J.~E.}\
  \bibnamefont {Pearson}}, \bibinfo {author} {\bibfnamefont {X.}~\bibnamefont
  {Cheng}}, \bibinfo {author} {\bibfnamefont {O.}~\bibnamefont {Heinonen}},
  \bibinfo {author} {\bibfnamefont {K.~L.}\ \bibnamefont {Wang}}, \bibinfo
  {author} {\bibfnamefont {Y.}~\bibnamefont {Zhou}}, \bibinfo {author}
  {\bibfnamefont {A.}~\bibnamefont {Hoffmann}}, \bibinfo {author}
  {\bibfnamefont {S.}~\bibnamefont {te~Velthuis}}, \ and\ \bibinfo {author}
  {\bibfnamefont {G.}~\bibnamefont {E.}},\ }\href {\doibase 10.1038/nphys3883}
  {\bibfield  {journal} {\bibinfo  {journal} {Nature Physics}\ }\textbf
  {\bibinfo {volume} {13}},\ \bibinfo {pages} {162} (\bibinfo {year}
  {2016})}\BibitemShut {NoStop}%
\bibitem [{\citenamefont {Legrand}\ \emph {et~al.}(2017)\citenamefont
  {Legrand}, \citenamefont {Maccariello}, \citenamefont {Reyren}, \citenamefont
  {Garcia}, \citenamefont {Moutafis}, \citenamefont {Moreau-Luchaire},
  \citenamefont {Collin}, \citenamefont {Bouzehouane}, \citenamefont {Cros},\
  and\ \citenamefont {Fert}}]{Legrand2017}%
  \BibitemOpen
  \bibfield  {author} {\bibinfo {author} {\bibfnamefont {W.}~\bibnamefont
  {Legrand}}, \bibinfo {author} {\bibfnamefont {D.}~\bibnamefont
  {Maccariello}}, \bibinfo {author} {\bibfnamefont {N.}~\bibnamefont {Reyren}},
  \bibinfo {author} {\bibfnamefont {K.}~\bibnamefont {Garcia}}, \bibinfo
  {author} {\bibfnamefont {C.}~\bibnamefont {Moutafis}}, \bibinfo {author}
  {\bibfnamefont {C.}~\bibnamefont {Moreau-Luchaire}}, \bibinfo {author}
  {\bibfnamefont {S.}~\bibnamefont {Collin}}, \bibinfo {author} {\bibfnamefont
  {K.}~\bibnamefont {Bouzehouane}}, \bibinfo {author} {\bibfnamefont
  {V.}~\bibnamefont {Cros}}, \ and\ \bibinfo {author} {\bibfnamefont
  {A.}~\bibnamefont {Fert}},\ }\href {\doibase 10.1021/acs.nanolett.7b00649}
  {\bibfield  {journal} {\bibinfo  {journal} {Nano Letters}\ }\textbf {\bibinfo
  {volume} {17}},\ \bibinfo {pages} {2703} (\bibinfo {year}
  {2017})}\BibitemShut {NoStop}%
\bibitem [{\citenamefont {Moreau-Luchaire}\ \emph {et~al.}(2016)\citenamefont
  {Moreau-Luchaire}, \citenamefont {Moutafis}, \citenamefont {Reyren},
  \citenamefont {Sampaio}, \citenamefont {Vaz}, \citenamefont {{Van Horne}},
  \citenamefont {Bouzehouane}, \citenamefont {Garcia}, \citenamefont
  {Deranlot}, \citenamefont {Warnicke}, \citenamefont {Wohlh{\"{u}}ter},
  \citenamefont {George}, \citenamefont {Weigand}, \citenamefont {Raabe},
  \citenamefont {Cros},\ and\ \citenamefont {Fert}}]{Moreau-Luchaire2016}%
  \BibitemOpen
  \bibfield  {author} {\bibinfo {author} {\bibfnamefont {C.}~\bibnamefont
  {Moreau-Luchaire}}, \bibinfo {author} {\bibfnamefont {C.}~\bibnamefont
  {Moutafis}}, \bibinfo {author} {\bibfnamefont {N.}~\bibnamefont {Reyren}},
  \bibinfo {author} {\bibfnamefont {J.}~\bibnamefont {Sampaio}}, \bibinfo
  {author} {\bibfnamefont {C.~A.~F.}\ \bibnamefont {Vaz}}, \bibinfo {author}
  {\bibfnamefont {N.}~\bibnamefont {{Van Horne}}}, \bibinfo {author}
  {\bibfnamefont {K.}~\bibnamefont {Bouzehouane}}, \bibinfo {author}
  {\bibfnamefont {K.}~\bibnamefont {Garcia}}, \bibinfo {author} {\bibfnamefont
  {C.}~\bibnamefont {Deranlot}}, \bibinfo {author} {\bibfnamefont
  {P.}~\bibnamefont {Warnicke}}, \bibinfo {author} {\bibfnamefont
  {P.}~\bibnamefont {Wohlh{\"{u}}ter}}, \bibinfo {author} {\bibfnamefont
  {J.-M.}\ \bibnamefont {George}}, \bibinfo {author} {\bibfnamefont
  {M.}~\bibnamefont {Weigand}}, \bibinfo {author} {\bibfnamefont
  {J.}~\bibnamefont {Raabe}}, \bibinfo {author} {\bibfnamefont
  {V.}~\bibnamefont {Cros}}, \ and\ \bibinfo {author} {\bibfnamefont
  {A.}~\bibnamefont {Fert}},\ }\href {\doibase 10.1038/nnano.2015.313}
  {\bibfield  {journal} {\bibinfo  {journal} {Nature Nanotechnology}\ }\textbf
  {\bibinfo {volume} {11}},\ \bibinfo {pages} {444} (\bibinfo {year}
  {2016})}\BibitemShut {NoStop}%
\bibitem [{\citenamefont {Boulle}\ \emph {et~al.}(2016)\citenamefont {Boulle},
  \citenamefont {Vogel}, \citenamefont {Yang}, \citenamefont {Pizzini},
  \citenamefont {{de Souza Chaves}}, \citenamefont {Locatelli}, \citenamefont
  {Menteş}, \citenamefont {Sala}, \citenamefont {Buda-Prejbeanu},
  \citenamefont {Klein}, \citenamefont {Belmeguenai}, \citenamefont
  {Roussign{\'{e}}}, \citenamefont {Stashkevich}, \citenamefont {Ch{\'{e}}rif},
  \citenamefont {Aballe}, \citenamefont {Foerster}, \citenamefont {Chshiev},
  \citenamefont {Auffret}, \citenamefont {Miron},\ and\ \citenamefont
  {Gaudin}}]{Boulle2016}%
  \BibitemOpen
  \bibfield  {author} {\bibinfo {author} {\bibfnamefont {O.}~\bibnamefont
  {Boulle}}, \bibinfo {author} {\bibfnamefont {J.}~\bibnamefont {Vogel}},
  \bibinfo {author} {\bibfnamefont {H.}~\bibnamefont {Yang}}, \bibinfo {author}
  {\bibfnamefont {S.}~\bibnamefont {Pizzini}}, \bibinfo {author} {\bibfnamefont
  {D.}~\bibnamefont {{de Souza Chaves}}}, \bibinfo {author} {\bibfnamefont
  {A.}~\bibnamefont {Locatelli}}, \bibinfo {author} {\bibfnamefont {T.~O.}\
  \bibnamefont {Menteş}}, \bibinfo {author} {\bibfnamefont {A.}~\bibnamefont
  {Sala}}, \bibinfo {author} {\bibfnamefont {L.~D.}\ \bibnamefont
  {Buda-Prejbeanu}}, \bibinfo {author} {\bibfnamefont {O.}~\bibnamefont
  {Klein}}, \bibinfo {author} {\bibfnamefont {M.}~\bibnamefont {Belmeguenai}},
  \bibinfo {author} {\bibfnamefont {Y.}~\bibnamefont {Roussign{\'{e}}}},
  \bibinfo {author} {\bibfnamefont {A.}~\bibnamefont {Stashkevich}}, \bibinfo
  {author} {\bibfnamefont {S.~M.}\ \bibnamefont {Ch{\'{e}}rif}}, \bibinfo
  {author} {\bibfnamefont {L.}~\bibnamefont {Aballe}}, \bibinfo {author}
  {\bibfnamefont {M.}~\bibnamefont {Foerster}}, \bibinfo {author}
  {\bibfnamefont {M.}~\bibnamefont {Chshiev}}, \bibinfo {author} {\bibfnamefont
  {S.}~\bibnamefont {Auffret}}, \bibinfo {author} {\bibfnamefont {I.~M.}\
  \bibnamefont {Miron}}, \ and\ \bibinfo {author} {\bibfnamefont
  {G.}~\bibnamefont {Gaudin}},\ }\href {\doibase 10.1038/nnano.2015.315}
  {\bibfield  {journal} {\bibinfo  {journal} {Nature Nanotechnology}\ }\textbf
  {\bibinfo {volume} {11}},\ \bibinfo {pages} {449} (\bibinfo {year}
  {2016})}\BibitemShut {NoStop}%
\bibitem [{\citenamefont {Ma}\ \emph {et~al.}(2016)\citenamefont {Ma},
  \citenamefont {Yu}, \citenamefont {Li}, \citenamefont {Wang}, \citenamefont
  {Wu}, \citenamefont {Olsson}, \citenamefont {Chu}, \citenamefont {An},
  \citenamefont {Xiao}, \citenamefont {Wang},\ and\ \citenamefont
  {Li}}]{Ma2016}%
  \BibitemOpen
  \bibfield  {author} {\bibinfo {author} {\bibfnamefont {X.}~\bibnamefont
  {Ma}}, \bibinfo {author} {\bibfnamefont {G.}~\bibnamefont {Yu}}, \bibinfo
  {author} {\bibfnamefont {X.}~\bibnamefont {Li}}, \bibinfo {author}
  {\bibfnamefont {T.}~\bibnamefont {Wang}}, \bibinfo {author} {\bibfnamefont
  {D.}~\bibnamefont {Wu}}, \bibinfo {author} {\bibfnamefont {K.~S.}\
  \bibnamefont {Olsson}}, \bibinfo {author} {\bibfnamefont {Z.}~\bibnamefont
  {Chu}}, \bibinfo {author} {\bibfnamefont {K.}~\bibnamefont {An}}, \bibinfo
  {author} {\bibfnamefont {J.~Q.}\ \bibnamefont {Xiao}}, \bibinfo {author}
  {\bibfnamefont {K.~L.}\ \bibnamefont {Wang}}, \ and\ \bibinfo {author}
  {\bibfnamefont {X.}~\bibnamefont {Li}},\ }\href {\doibase
  10.1103/PhysRevB.94.180408} {\bibfield  {journal} {\bibinfo  {journal}
  {Physical Review B}\ }\textbf {\bibinfo {volume} {94}},\ \bibinfo {pages}
  {180408} (\bibinfo {year} {2016})}\BibitemShut {NoStop}%
\bibitem [{\citenamefont {Hsu}\ \emph {et~al.}(2018)\citenamefont {Hsu},
  \citenamefont {R{\'{o}}zsa}, \citenamefont {Finco}, \citenamefont {Schmidt},
  \citenamefont {Palot{\'{a}}s}, \citenamefont {Vedmedenko}, \citenamefont
  {Udvardi}, \citenamefont {Szunyogh}, \citenamefont {Kubetzka}, \citenamefont
  {von Bergmann},\ and\ \citenamefont {Wiesendanger}}]{Hsu2018}%
  \BibitemOpen
  \bibfield  {author} {\bibinfo {author} {\bibfnamefont {P.-J.}\ \bibnamefont
  {Hsu}}, \bibinfo {author} {\bibfnamefont {L.}~\bibnamefont {R{\'{o}}zsa}},
  \bibinfo {author} {\bibfnamefont {A.}~\bibnamefont {Finco}}, \bibinfo
  {author} {\bibfnamefont {L.}~\bibnamefont {Schmidt}}, \bibinfo {author}
  {\bibfnamefont {K.}~\bibnamefont {Palot{\'{a}}s}}, \bibinfo {author}
  {\bibfnamefont {E.}~\bibnamefont {Vedmedenko}}, \bibinfo {author}
  {\bibfnamefont {L.}~\bibnamefont {Udvardi}}, \bibinfo {author} {\bibfnamefont
  {L.}~\bibnamefont {Szunyogh}}, \bibinfo {author} {\bibfnamefont
  {A.}~\bibnamefont {Kubetzka}}, \bibinfo {author} {\bibfnamefont
  {K.}~\bibnamefont {von Bergmann}}, \ and\ \bibinfo {author} {\bibfnamefont
  {R.}~\bibnamefont {Wiesendanger}},\ }\href {\doibase
  10.1038/s41467-018-04015-z} {\bibfield  {journal} {\bibinfo  {journal}
  {Nature Communications}\ }\textbf {\bibinfo {volume} {9}},\ \bibinfo {pages}
  {1571} (\bibinfo {year} {2018})}\BibitemShut {NoStop}%
\bibitem [{\citenamefont {Srivastava}\ \emph {et~al.}(2018)\citenamefont
  {Srivastava}, \citenamefont {Schott}, \citenamefont {Juge}, \citenamefont
  {Kři{\v{z}}{\'{a}}kov{\'{a}}}, \citenamefont {Belmeguenai}, \citenamefont
  {Roussign{\'{e}}}, \citenamefont {Bernand-Mantel}, \citenamefont {Ranno},
  \citenamefont {Pizzini}, \citenamefont {Ch{\'{e}}rif}, \citenamefont
  {Stashkevich}, \citenamefont {Auffret}, \citenamefont {Boulle}, \citenamefont
  {Gaudin}, \citenamefont {Chshiev}, \citenamefont {Baraduc},\ and\
  \citenamefont {B{\'{e}}a}}]{Srivastava2018}%
  \BibitemOpen
  \bibfield  {author} {\bibinfo {author} {\bibfnamefont {T.}~\bibnamefont
  {Srivastava}}, \bibinfo {author} {\bibfnamefont {M.}~\bibnamefont {Schott}},
  \bibinfo {author} {\bibfnamefont {R.}~\bibnamefont {Juge}}, \bibinfo {author}
  {\bibfnamefont {V.}~\bibnamefont {Kři{\v{z}}{\'{a}}kov{\'{a}}}}, \bibinfo
  {author} {\bibfnamefont {M.}~\bibnamefont {Belmeguenai}}, \bibinfo {author}
  {\bibfnamefont {Y.}~\bibnamefont {Roussign{\'{e}}}}, \bibinfo {author}
  {\bibfnamefont {A.}~\bibnamefont {Bernand-Mantel}}, \bibinfo {author}
  {\bibfnamefont {L.}~\bibnamefont {Ranno}}, \bibinfo {author} {\bibfnamefont
  {S.}~\bibnamefont {Pizzini}}, \bibinfo {author} {\bibfnamefont {S.-M.}\
  \bibnamefont {Ch{\'{e}}rif}}, \bibinfo {author} {\bibfnamefont
  {A.}~\bibnamefont {Stashkevich}}, \bibinfo {author} {\bibfnamefont
  {S.}~\bibnamefont {Auffret}}, \bibinfo {author} {\bibfnamefont
  {O.}~\bibnamefont {Boulle}}, \bibinfo {author} {\bibfnamefont
  {G.}~\bibnamefont {Gaudin}}, \bibinfo {author} {\bibfnamefont
  {M.}~\bibnamefont {Chshiev}}, \bibinfo {author} {\bibfnamefont
  {C.}~\bibnamefont {Baraduc}}, \ and\ \bibinfo {author} {\bibfnamefont
  {H.}~\bibnamefont {B{\'{e}}a}},\ }\href {\doibase
  10.1021/acs.nanolett.8b01502} {\bibfield  {journal} {\bibinfo  {journal}
  {Nano Letters}\ ,\ \bibinfo {pages} {acs.nanolett.8b01502}} (\bibinfo {year}
  {2018})}\BibitemShut {NoStop}%
\bibitem [{\citenamefont {Romming}\ \emph {et~al.}(2013)\citenamefont
  {Romming}, \citenamefont {Hanneken}, \citenamefont {Menzel}, \citenamefont
  {Bickel}, \citenamefont {Wolter}, \citenamefont {von Bergmann}, \citenamefont
  {Kubetzka},\ and\ \citenamefont {Wiesendanger}}]{Romming2013}%
  \BibitemOpen
  \bibfield  {author} {\bibinfo {author} {\bibfnamefont {N.}~\bibnamefont
  {Romming}}, \bibinfo {author} {\bibfnamefont {C.}~\bibnamefont {Hanneken}},
  \bibinfo {author} {\bibfnamefont {M.}~\bibnamefont {Menzel}}, \bibinfo
  {author} {\bibfnamefont {J.~E.}\ \bibnamefont {Bickel}}, \bibinfo {author}
  {\bibfnamefont {B.}~\bibnamefont {Wolter}}, \bibinfo {author} {\bibfnamefont
  {K.}~\bibnamefont {von Bergmann}}, \bibinfo {author} {\bibfnamefont
  {A.}~\bibnamefont {Kubetzka}}, \ and\ \bibinfo {author} {\bibfnamefont
  {R.}~\bibnamefont {Wiesendanger}},\ }\href {\doibase 10.1126/science.1240573}
  {\bibfield  {journal} {\bibinfo  {journal} {Science (New York, N.Y.)}\
  }\textbf {\bibinfo {volume} {341}},\ \bibinfo {pages} {636} (\bibinfo {year}
  {2013})}\BibitemShut {NoStop}%
\bibitem [{\citenamefont {Balk}\ \emph {et~al.}(2017)\citenamefont {Balk},
  \citenamefont {Kim}, \citenamefont {Pierce}, \citenamefont {Stiles},
  \citenamefont {Unguris},\ and\ \citenamefont {Stavis}}]{Balk2017}%
  \BibitemOpen
  \bibfield  {author} {\bibinfo {author} {\bibfnamefont {A.}~\bibnamefont
  {Balk}}, \bibinfo {author} {\bibfnamefont {K.-W.}\ \bibnamefont {Kim}},
  \bibinfo {author} {\bibfnamefont {D.}~\bibnamefont {Pierce}}, \bibinfo
  {author} {\bibfnamefont {M.}~\bibnamefont {Stiles}}, \bibinfo {author}
  {\bibfnamefont {J.}~\bibnamefont {Unguris}}, \ and\ \bibinfo {author}
  {\bibfnamefont {S.}~\bibnamefont {Stavis}},\ }\href {\doibase
  10.1103/PhysRevLett.119.077205} {\bibfield  {journal} {\bibinfo  {journal}
  {Physical Review Letters}\ }\textbf {\bibinfo {volume} {119}},\ \bibinfo
  {pages} {077205} (\bibinfo {year} {2017})}\BibitemShut {NoStop}%
\bibitem [{\citenamefont {Wells}\ \emph {et~al.}(2017)\citenamefont {Wells},
  \citenamefont {Shepley}, \citenamefont {Marrows},\ and\ \citenamefont
  {Moore}}]{Wells2017}%
  \BibitemOpen
  \bibfield  {author} {\bibinfo {author} {\bibfnamefont {A.~W.~J.}\
  \bibnamefont {Wells}}, \bibinfo {author} {\bibfnamefont {P.~M.}\ \bibnamefont
  {Shepley}}, \bibinfo {author} {\bibfnamefont {C.~H.}\ \bibnamefont
  {Marrows}}, \ and\ \bibinfo {author} {\bibfnamefont {T.~A.}\ \bibnamefont
  {Moore}},\ }\href {\doibase 10.1103/PhysRevB.95.054428} {\bibfield  {journal}
  {\bibinfo  {journal} {Physical Review B}\ }\textbf {\bibinfo {volume} {95}},\
  \bibinfo {pages} {054428} (\bibinfo {year} {2017})}\BibitemShut {NoStop}%
\bibitem [{\citenamefont {Ivanov}\ and\ \citenamefont
  {Zaspel}(2007)}]{Ivanov2007}%
  \BibitemOpen
  \bibfield  {author} {\bibinfo {author} {\bibfnamefont {B.~A.}\ \bibnamefont
  {Ivanov}}\ and\ \bibinfo {author} {\bibfnamefont {C.~E.}\ \bibnamefont
  {Zaspel}},\ }\href {\doibase 10.1103/PhysRevLett.99.247208} {\bibfield
  {journal} {\bibinfo  {journal} {Physical Review Letters}\ }\textbf {\bibinfo
  {volume} {99}},\ \bibinfo {pages} {247208} (\bibinfo {year}
  {2007})}\BibitemShut {NoStop}%
\bibitem [{\citenamefont {Liu}\ \emph {et~al.}(2007)\citenamefont {Liu},
  \citenamefont {He},\ and\ \citenamefont {Zhang}}]{Liu2007}%
  \BibitemOpen
  \bibfield  {author} {\bibinfo {author} {\bibfnamefont {Y.}~\bibnamefont
  {Liu}}, \bibinfo {author} {\bibfnamefont {H.}~\bibnamefont {He}}, \ and\
  \bibinfo {author} {\bibfnamefont {Z.}~\bibnamefont {Zhang}},\ }\href
  {\doibase 10.1063/1.2822436} {\bibfield  {journal} {\bibinfo  {journal}
  {Applied Physics Letters}\ }\textbf {\bibinfo {volume} {91}},\ \bibinfo
  {pages} {242501} (\bibinfo {year} {2007})}\BibitemShut {NoStop}%
\bibitem [{\citenamefont {He}\ \emph {et~al.}(2017)\citenamefont {He},
  \citenamefont {Angizi},\ and\ \citenamefont {Fan}}]{He2017}%
  \BibitemOpen
  \bibfield  {author} {\bibinfo {author} {\bibfnamefont {Z.}~\bibnamefont
  {He}}, \bibinfo {author} {\bibfnamefont {S.}~\bibnamefont {Angizi}}, \ and\
  \bibinfo {author} {\bibfnamefont {D.}~\bibnamefont {Fan}},\ }\href {\doibase
  10.1109/LMAG.2017.2689721} {\bibfield  {journal} {\bibinfo  {journal} {IEEE
  Magnetics Letters}\ }\textbf {\bibinfo {volume} {8}},\ \bibinfo {pages} {1}
  (\bibinfo {year} {2017})}\BibitemShut {NoStop}%
\end{thebibliography}%

\end{document}